\documentclass[%
 reprint,
 amsmath,amssymb,
 aps,
]{revtex4-1}

\usepackage{graphicx}% Include figure files
\usepackage{dcolumn}% Align table columns on decimal point
\usepackage{bm}% bold math
\usepackage{subfig}% bold math
\begin{document}

\preprint{}

\title{ Dynamic multi-factor clustering of financial networks}
\author{Gordon J Ross}
\affiliation{%
Heilbronn Institute for Mathematical Research, University of Bristol, United Kingdom\\
gordon.ross@bristol.ac.uk 
}%
\date{\today}% It is always \today, today,
             %  but any date may be explicitly specified

\begin{abstract}
We investigate the tendency for financial instruments to form clusters when there are multiple factors influencing the correlation structure. Specifically, we consider a stock portfolio which contains companies from different industrial sectors, located in several different countries. Both sector membership and geography combine to create a complex clustering structure where companies seem to first be divided based on sector, with geographical subclusters emerging within each industrial sector. We argue that standard techniques for detecting overlapping clusters and communities are not able to capture this type of structure, and show how robust regression techniques can instead be used to remove the influence of both sector and geography from the correlation matrix separately. %These adjusted returns display a cleaner clustering structure, which allows us to study its behaviour over time. 
Our analysis reveals that prior to the 2008 financial crisis, companies did not tend to form clusters based on geography. This changed immediately following the crisis, with geography becoming a more important determinant of clustering.% This has practical implications for risk and portfolio management techniques, which must take into account potential changes in the underlying factors which determine the observed correlation structure.

\end{abstract}

\pacs{Valid PACS appear here}% PACS, the Physics and Astronomy
                             % Classification Scheme.
%\keywords{Suggested keywords}%Use showkeys class option if keyword
                              %display desired
\maketitle

%\tableofcontents

%rewrite abstract to make more general
\section{Introduction}

%Complex networks arise in many diverse scientific fields such as protein-protein interaction networks in biology \cite{Junker2008}, academic citation networks \cite{Hopcroft2004}, and the study of social interactions \cite{Heard2010}. It is now widely accepted that traditional random  Erdos-Renyi random graph models are inadequate for modeling the features which can be found in many real networks such as power law degree distributions  \cite{Reka2002},  and the presence of clustering patterns where the network seems to be partitioned into subgroups of nodes  exhibiting similar behavior \cite{Fortunato2010}. Recent interdisciplinary work has focused both on developing  models which more accurately capture the structures found in complex networks, and on analyzing empirical networks in order to better understand and characterize the features which they possess. 

Financial markets provide a rich application area  for the study of complex networks \cite{Mantegna1999a,Coronnello2005,Song2011,Fenn2011,Bonanno2003}. In a financial network, nodes represent financial instruments while the edge weights denote the correlation between the price changes of these instruments over some period of time. %For the remainder of the paper we assume that we are dealing with the stock prices of individual companies, although the methods discussed are more general than this and apply equally to networks based on currency exchange rates, commodity prices, and so on.
The purpose of studying financial networks is to gain information about the underlying correlation structure and to better understand how it is influenced by various real-world factors, which is of great importance to risk management  \cite{Jorion2006}, investment theory \cite{Elton2006}, and index tracking  \cite{Coriellia2006}.

%. For example, in risk management it is desirable to be able to measure the risk associated with a portfolio of assets \cite{Jorion2006}, which is a function of their joint correlation matrix. This is an important task since if too many assets in a portfolio are highly correlated, then the portfolio is in danger of experiencing large price changes in response to real world events which influence all companies simultaneously. Correlation analysis is also important in other areas of finance such as investment theory \cite{Elton2006}, and index tracking  \cite{Coriellia2006}.

Much work has focused  on the task of detecting clusters in financial networks, corresponding to groups of companies which have highly correlated stock price movements. %Despite early work suggesting that much of the observed empirical correlation  between pairs of companies is due to random noise \cite{Laloux1999}, 
It has been found that companies often form clusters based on the industrial sector to which they belong, with (for example) banks tending to behave differently than telecommunications companies \cite{ Garas2007,Bonanno2003,Kantar2011,Tumminello2007}. These studies generally only focus on the analysis of companies located within a single country,  however a similar analysis has been carried out  currency exchange rates and national stock indexes \cite{Fenn2009,Eryigita2009,Keskin2011}  which shows that there is also evidence of clustering based on geographical location, with East Asian currencies tending to behave differently than European ones, and so on.

The existing literature on financial correlation networks has mainly focused on situations where there is only a single factor which influences the correlation structure -   membership when studying individual companies, and geographical location when studying currency exchange rates or national stock indexes. However, many real portfolios will be driven by multiple factors. For example, a portfolio containing companies which are spread across multiple sectors and  located in several different countries may be expected to form clusters which are based on both sector and geography, with Spanish banks perhaps behaving in a different manner than  British banks. %such as the Euro Stoxx 50 index, the Standard and Poor's Europe 350 or the various iTraxx credit indexes. 
An analysis which focused only on one clustering factor, such as sector, would be inadequate for this type of data. With this in mind, we develop techniques which allow multiple factors to drive the correlation structure. We show that in a naive analysis, it appears that it is mainly sector membership which influences clustering. However a closer look reveals that there is a semblance of geographical clustering, although this is swamped by sector effects. In order to reveal the true influence of geography, we can make a transformation which strips the correlation matrix of the influence of sector membership. % Our approach is quite general and would apply to non-financial networks which are similarly driven by multiple factors.

Finally, we investigate how the influence of sector and geography  on the clustering structure evolves over time, with a particularly important change occurring during the recent financial crisis where geography starts to become a more important determinant in the countries which were most affected by the sovereign debt crisis. Understanding the time-dynamics of financial correlations is an important area of research \cite{Onnela2003, Conlon2009,Fenn2011}, and our analysis shows how this can sensibly be carried out in cases where there are multiple latent clustering factors.

%We investigate this time-varying aspect of the cluster structure. This is closely related to recent work on evolving community detection in graphs such as that of \cite{Palla2007, Hopcroft2004}. However our focus is to investigate how sector and geography interact, and we propose a test to determine which is more influential at a given point in time. To give a preview of our results, we find that under normal circumstances, sector membership is the most influential factor in determining the correlation structure, although geography does play a significant role. However during the recent financial crisis, the influence of geography increases, particular for companies in regions such as Spain and Italy which have experienced particularly severe crises.

%THIS SECTION SI FINE
\section{Data}
\label{sec:data}
%daily vs weekly
The data we will analyse throughout consists of the daily returns of the $350$ companies which compose  the Standard \& Poor (S\&P) 350 European stock index. The companies in this index are among the largest in the continent, and together make up approximately $70\%$ of the total European market capitalization.

We obtained the daily closing prices for each company between the dates of January 1$^{st}$ 2003, and March 17$^{th}$ 2012. In order to avoid the difficult problem of synchronizing prices across substantially different time zones, we have excluded companies which trade only on non-European stock exchanges, along with those which either did not have stock quotes going back to $1$ January $2003$, or which underwent mergers or acquisitions during this period. In total, there are $267$ companies remaining for which we have data. In order to avoid issues related to the effects of different currencies, we converted the closing prices of each company into Euros by using the relevant historical exchange rate.

Finally, Standard and Poor provide an official classification of these companies into industry sectors based on the Global Industry Classification Standard (GICS), with each company being assigned to one of  $10$ different possible sectors. We will use this in order to check whether the clusters found in the data reproduce real patterns.The number of companies in the data set belonging to each sector and geographical region is shown in Table \ref{tab:companysectorlist}, where we have omitted countries containing less than $5$ companies since such small samples may not allow for accurate analysis. It is important to note that the GICS sector classifications are very broad and many subdivisions exist within each one; for example, the group labelled as Financials includes both banks and insurance companies which may behave in different ways, while the oil companies within the Energy sector may behave slightly different from other  non-oil energy companies. This point will be important when it comes to assessing the quality of clustering structures in Sec. \ref{sec:purity}.
%need to explain how broad classification is

\begin{table}[t]
\begin{center}
\begin{tabular}{lrlr}
  \hline
Sector & Number & Country & Number \\ 
  \hline
Materials & 30 & Great Britain & 79 \\ 
  Consumer Staples & 24 & France & 44 \\ 
  Financials & 57 & Germany & 33 \\ 
  Utilities & 16 & Sweden & 24 \\ 
  Telecommunications Services & 15 & Italy & 18 \\ 
  Industrials & 45 & Spain & 16 \\ 
  Consumer Discretionary & 44 & Switzerland & 12 \\ 
  Information Technology & 10 & Netherlands & 8 \\ 
  Energy & 11 & Greece & 5 \\ 
  Health Care & 16 & & \\ 
   \hline
\end{tabular}
\caption{Number of companies from each sector and country present in the data set}
\label{tab:companysectorlist}
\end{center}
\end{table}

%this section is fine
\section{Clustering Algorithm}
\label{sec:clustering}
Clustering is the process of assigning objects to groups such that objects in the same group are similar according to some specified distance metric, while companies in distinct groups are different. As in  \cite{Mantegna1999a,Coronnello2005}., we put a distance metric on the space of companies which is based on the correlation between their daily stock log returns. Let $P_{i,t}$ denote the stock price in Euros of company $i$ on day $t$, and let $r_{i,t} = \log P_{i,t} - \log P_{i,t-1}$ denote the logarithm of the daily return. The Pearson correlation between companies $i$ and $j$ based over the whole period off $T$ equally weighted days is then:
\begin{equation}
\rho_{ij} =\frac{ \sum_{t=1}^{T} r_{i,t} \sum_{t=1}^{T} r_{j,t}} {\sqrt{\sum_{t=1}^{T} r_{i,t}^2 \sum_{t=1}^{T} r_{j,t}^2}}.
\label{eqn:correlation}
\end{equation}
The distance between any two companies $i$ and $j$ can then be defined as:
\begin{equation}
W_{ij} = \sqrt{2 (1-\rho_{ij})}.
\label{eqn:distancemetric}
\end{equation}
where it can be seen that $W_{ij} \in [0,1]$, with $W_{ij}=0$ if the companies are perfectly correlated.

Although there are many possible algorithms for clustering, we choose to use average link agglomerative hierarchal clustering (HC) \cite{Coronnello2005}. This is because as noted above the GICS sector labels are broad, and there is reason to believe that there may be a hierarchy of sub-sectors under each broad label. A full description of the HC algorithm can be found in the above reference, we will only summarise it here. Given the distance matrix $\{W_{ij}\}$,  HC performs clustering by recursively merging the companies into clusters based on greedy optimisation.  Initially each of the $n$ companies is assigned to its own cluster, resulting in the $n$ clusters $\{c_1,\ldots,c_n\}$ . Then, HC selects the pair of clusters which are closest together, and successively merges them until only one cluster remains. The output of the HC algorithm is a \textbf{dendrogram}. This is a tree structure which shows the order in which the clusters were merged. Each company is represented as a leaf of the tree. When two clusters are joined by HC, a link between them is added. The dendrogram allows the hierarchal clustering structure of the data to be visualized, since the relationship between companies and clusters and subclusters can be easily viewed.

\section{`Hidden' Clusters}

Before proceeding with the analysis of the data, we first describe the problems which can arise when multiple factors are influencing the clustering structure. In a typical stock portfolio, each company has an industrial sector membership  and also a geographical location. In times of normal market behavior,  we will later show that the clustering structure of the network is largely attributable to sector membership. However, there is a noticeable subclustering where for example banks located in the United Kingdom behave slightly different than banks located in Germany. This suggests that the structure of the correlation matrix is formed by the interaction of two distinct dimensions, one corresponding to sector membership and the other to geography. Because sector membership is usually the dominant factor influencing the correlation matrix, the effects of geography are somewhat masked in the raw data. 

\begin{figure}[]
  \centering
  \subfloat[]{\label{fig:intro1}\includegraphics[width=0.25\textwidth]{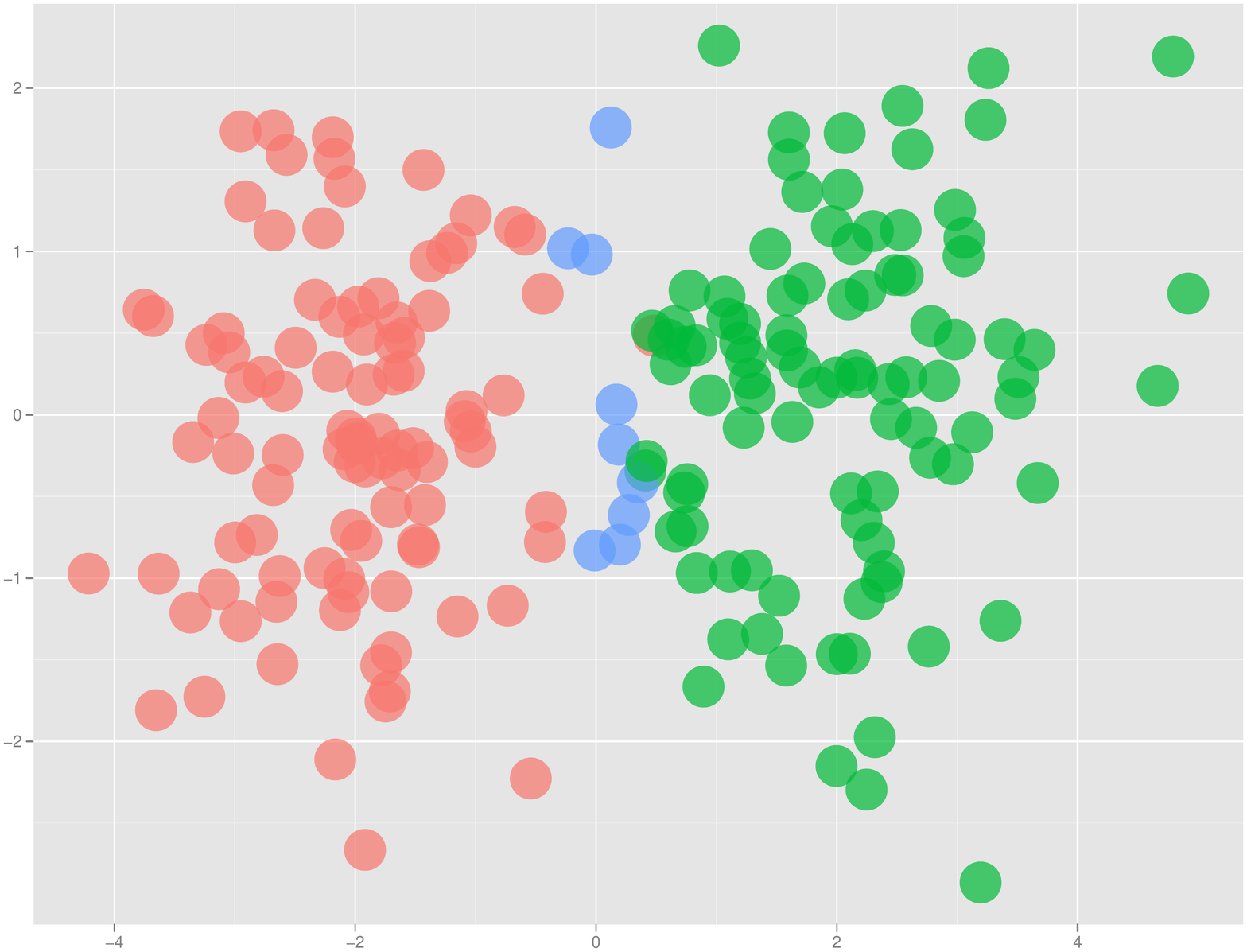}}
  \subfloat[]{\label{fig:intro2}\includegraphics[width=0.25\textwidth]{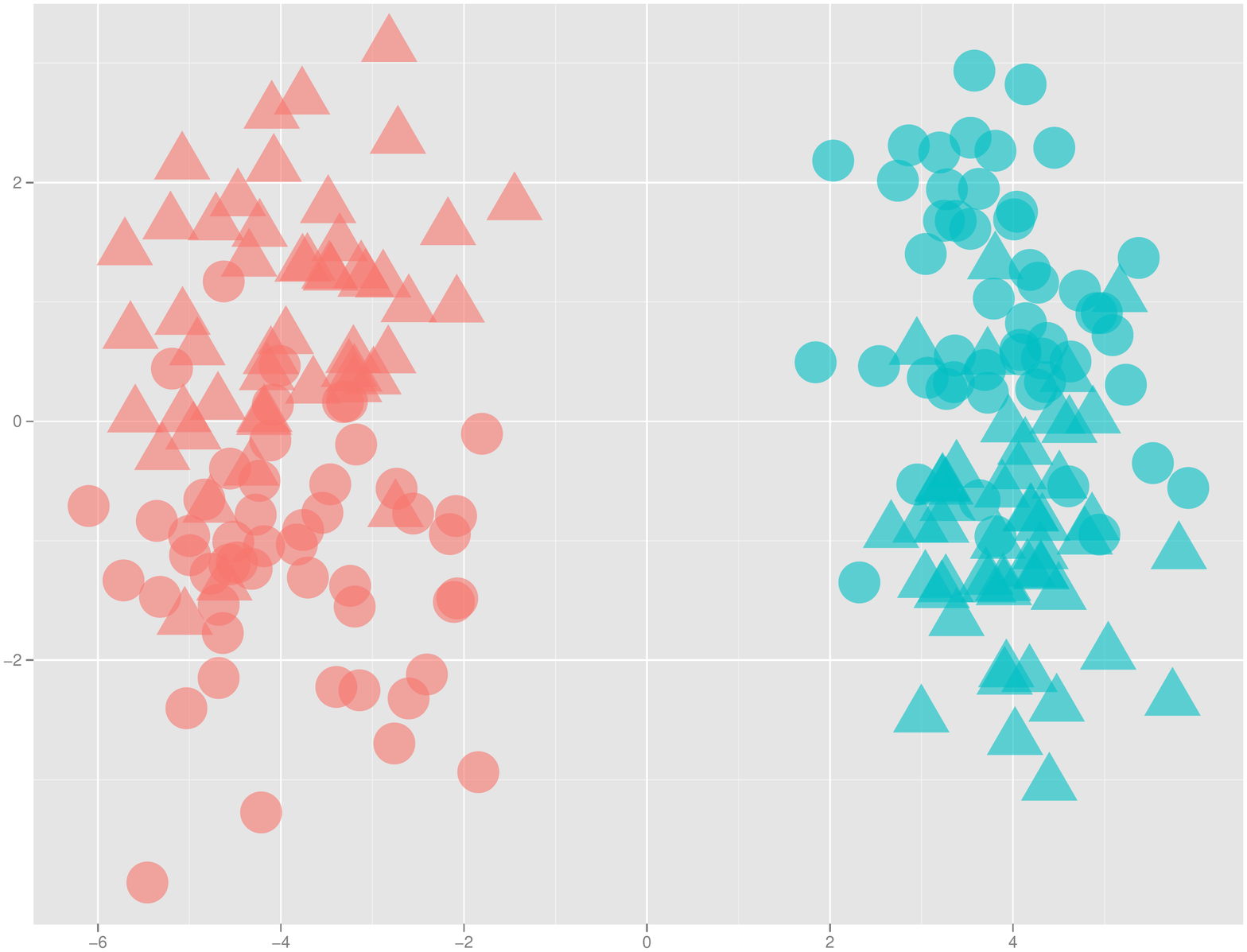}} 
  %ONE MORE SUBFLOAT SHOWING CLUSTERING ALGORIHTM OUTPUT
\caption{(a) shows a typical scenario where two clusters overlap, due to a number of points (colored in blue) belonging to both. (b) shows the 'hidden' cluster  situation where the effect of the feature which groups points as either circles or triangles is masked by the dominant feature based on color.}
  \label{fig:mds}
\end{figure}

This problem differs markedly from the usual scenario considered in the literature on detecting overlapping communities \cite{Palla2005} which assumes that, even though a node may belong to multiple communities, these different communities are distinct and well formed.  An example of this situation is shown in Fig. \ref{fig:intro1} which shows a representation of two overlapping clusters colored red and green, plotted in a 2 dimensional space. A small number of points belong to both clusters, and are plotted in blue  In contrast, our situation is akin to that shown in Fig. \ref{fig:intro2}. Here each point has two binary features attached to it, say A and B. If feature A has a value of 1 then the point is colored red, and if it has a value 0 then it is colored blue. Similarly if feature B has a value of 1 then the point is represented by a circle, otherwise it is represented by a triangle. From the plot, it is clear that feature A has the largest effect on the clustering structure, with points being very well separated based on which value they have. However feature B also has some weight, with points being sub-grouped within each cluster based on this. If we were to naively run a clustering algorithm on this data, then it would not be able to detect the effect that feature B has on the clustering because this is masked due to the stronger effect of A. In order to reveal the true effect that B is having, we must first transform the data to remove the effect of feature A in a principled manner.

A reasonable question which might be asked is: if the clustering structure implied by secondary factors such as geography are masked in the financial data to the extent that they are not immediately detectable, then why are they of interest? The simple answer is that the correlation structure of real data is usually not static, but instead changes over time. Although at one point in time the clustering based on geographical location may be masked, it may become more dominant in the future as we will show with reference to the 2008 financial crisis. %Indeed, we will show that during the post 2008 global economic crisis, geography became a significantly more important determinant of correlation. A naive analysis of the original correlation matrix would not have alerted one to this possibility, which may have drastic implications for financial trading strategies constructed under the assumption that the revealed sector-driven correlation structure was the only factor driving behavior.

%Although at one point in time the secondary factors may be masked, it is possible for them to become more dominant as time progresses. If their influence has not been detected, then one has no way of anticipating this which can lead to bad decision making. Again, financial data provides an excellent example of this phenomena; although sector membership does drive the correlation structure under normal circumstance, this can change in times of crisis. Through our analysis of the dynamic correlation structure of the network, we show that during the post 2008 global economic crisis, geography became a significantly more important determinant of correlation, with companies in particularly crisis-hit countries such as Spain and Greece starting to exhibit similar behavior. A naive analysis of the original correlation matrix would not have alerted one to this possibility, which may have drastic implications for financial trading strategies constructed under the assumption that the revealed sector-driven correlation structure was the only factor driving behavior.

%mention I'm using average link

\section{Static Analysis}
\label{sec:static}

We begin the data analysis by working with the static correlation matrix defined over the whole $8$ years of data, ignoring any time varying structure. In order to aid interpretation of the clusters found by the HC algorithm, it is useful to project the correlation matrix into a two-dimensional space in order to better visualize its structure. Although the actual clustering is performed in the original high-dimensional space rather than in two-dimensions, such a representation can provide an interesting visual summary of the data 

Multidimensional scaling (MDS) can be used for this purpose. MDS is a traditional tool in statistical analysis which allows high dimensional data to be plotted on the Euclidean plane \cite{Sibson1979}. Given a set of $n$ objects and an arbitrary distance matrix $\{W_ij\}$ where $W_{ij}$ denotes the distance between object $i$ and object $j$, MDS searches for a representation of the objects in two dimensions such that the distance between each pair of objects is as close as possible to the distance between them in the original higher dimensional space. In other words, it seeks a collection of $n$ vectors $x_i \in \mathbb{R}^2$ satisfying:

$$\min_{x_1,\ldots,x_n} \sum_{i < j} (||x_i - x_j|| - W_{ij})^2.$$

\begin{figure}[]
  \centering
  \subfloat[]{\label{fig:mdssector}\includegraphics[width=0.25\textwidth]{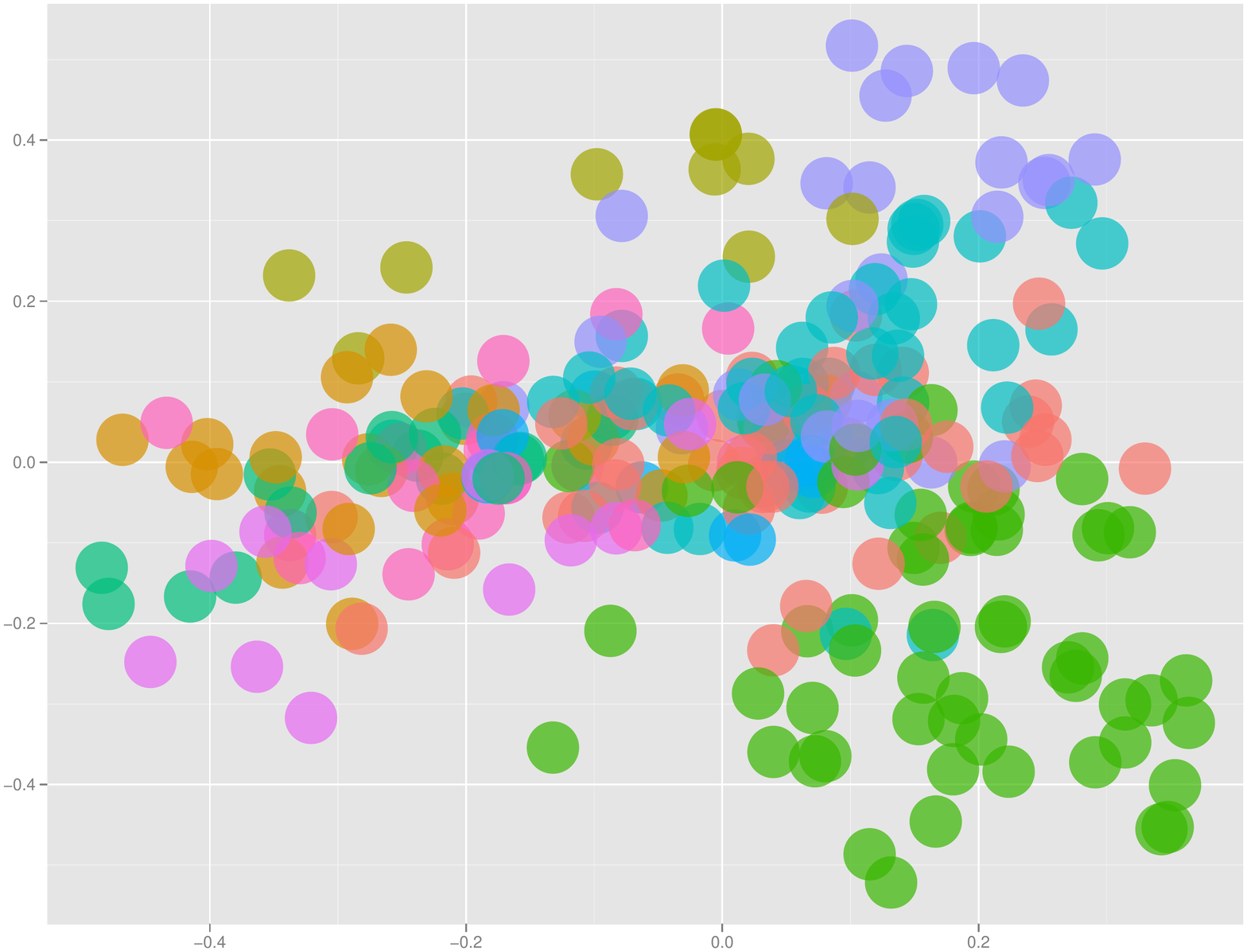}}
  \subfloat[  ]{\label{fig:mdsgeography}\includegraphics[width=0.25\textwidth]{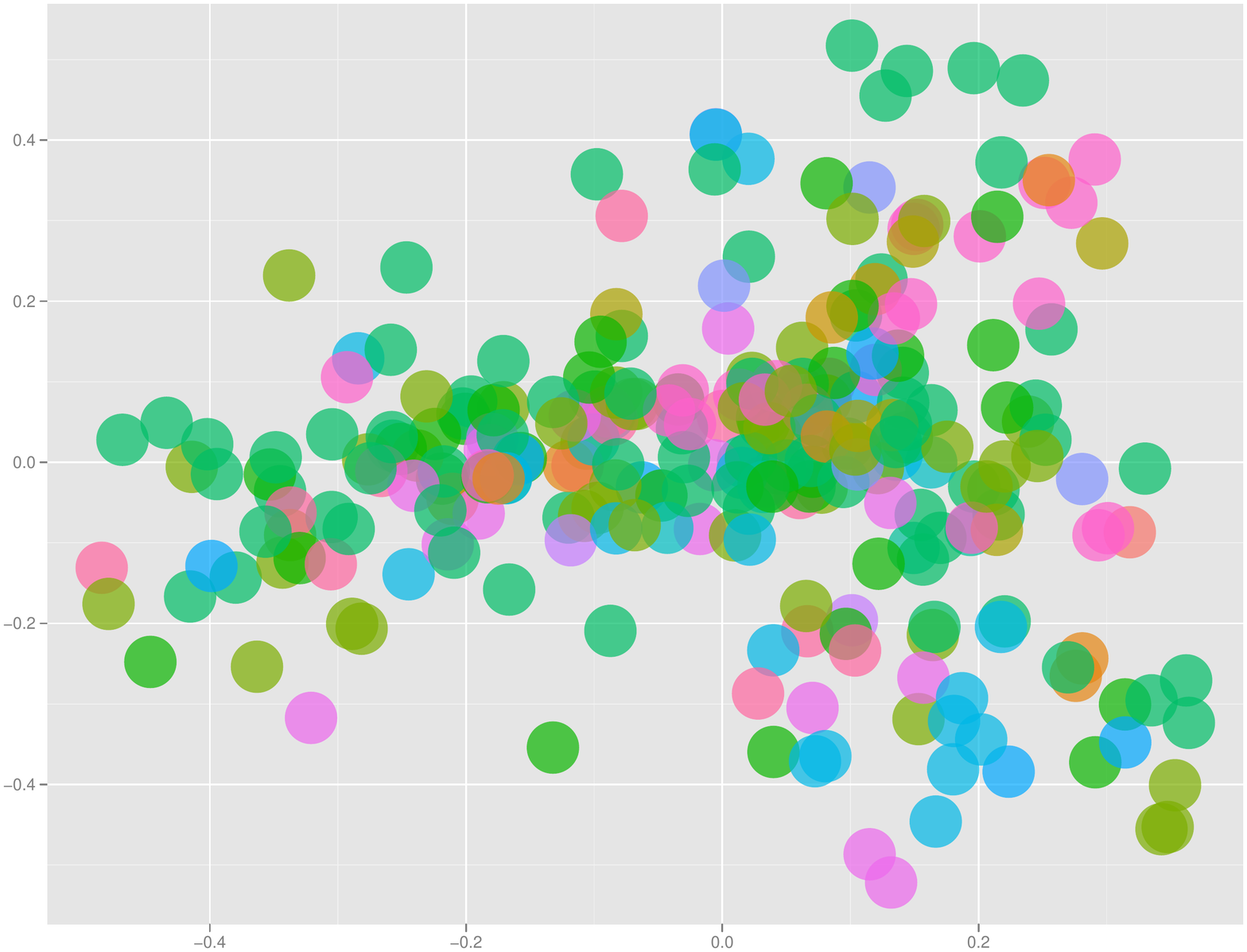}} \\        
  \subfloat[  ]{\label{fig:dendrosector}\includegraphics[width=0.25\textwidth]{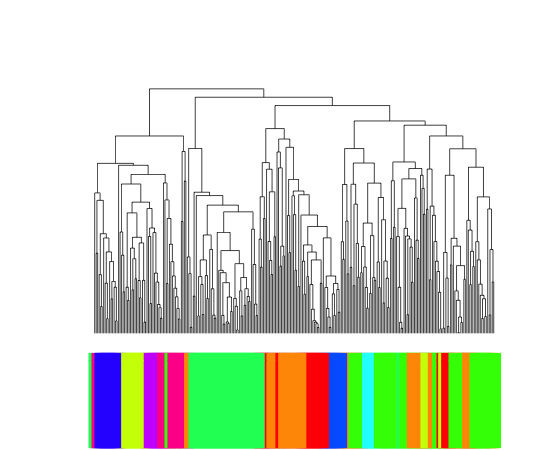}}
  \subfloat[  ]{\label{fig:dendrocountry}\includegraphics[width=0.25\textwidth]{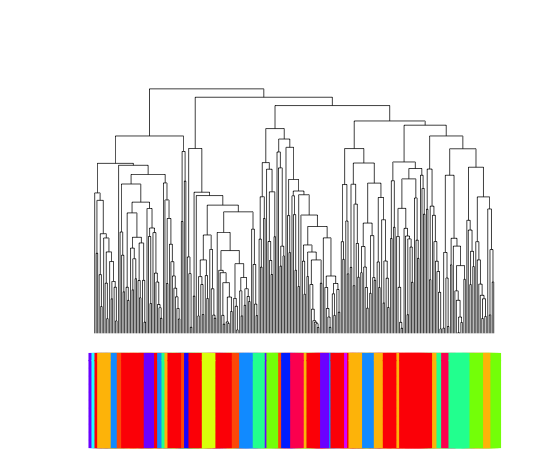}} 
\caption{Fig. (a) shows a MDS plot of the data with companies coloured based on their industrial sector membership. Fig. (b) shows the same, but coloured based on geographical location. Figs. (c) and (d) show the respective dendograms, with Fig. (c) coloured based on sector membership, and Fig. (d) coloured based on geographical location}
  \label{fig:mds2}
\end{figure}

%remember I used adjusted differences
Fig. \ref{fig:mdssector} shows a MDS plot of the data where each company is represented by a colored point denoting its sector, with each sector having a different color. If companies in the same sector do behave in a similar manner then they will be close to each other on this plot, and we would  see large groups of companies all of which have the same color. This is indeed what happens, with the companies in the Financials sector (colored green) in the bottom right corner being especially well separated from the rest. However, there is also substantial overlap between sectors; for example, looking at the top of the plot shows that while companies in the Industrials sector (blue) and Materials sector (cyan) companies do split off from the remainder of the companies, they tend to be quite similar to each other. This provides some indication of why previous approaches to clustering based on sector such as \cite{Coronnello2005}  have only achieved limited success; although there is sectoral grouping in the data, it is not strong enough for there to be a formation of completely disconnected sectors, and there is substantial overlap.

%%%fix descriptions here
Next, Fig.  \ref{fig:mdsgeography} shows the same MDS plot  with the companies instead colored based on their geographical location rather than their sector membership. It is visually obvious that there is less clustering structure than in the previous plot,  suggesting that geography does not have a strong influence on the clustering. However this naive analysis is slightly misleading. Consider the Industrial companies in the top right which were colored blue in the previous picture. Within these Industrials, there is a definite subclustering based on geography, which is most evident when looking at the group of five companies at the top of Fig. \ref{fig:mdsgeography}. These form a distinct group within the Industrial sector and have a different color from the rest of that sector, showing that they are in a different country from the rest (Germany). A similar picture can be found looking at the Financial companies in the bottom right of Fig. \ref{fig:mdsgeography}; again within this sector it can be seen that those in the same country tend to be closer together, suggesting  similar behavior. The same pattern can be observed throughout the plot, which suggests that the data has a structure similar to that discussed in the previous section where, although sector is the most important factor in determining the cluster structure of the data, geography also plays an important secondary  role when it comes to subclustering companies within each sector.. In other words, the companies seem to have a structure reminiscent of Fig. \ref{fig:intro2}.

\begin{table*}
\begin{center}
\subfloat{
\begin{tabular}{lrllrl}
\hline
\multicolumn{6}{c}{Before Transformation}\\
  \hline
Sector & Purity & & Country & Purity & \\ 
  \hline
Consumer Discretionary & 0.49 & ($< $ 0.001) & France & 0.22 & (0.004) \\ 
  Consumer Staples & 0.53 &  ($< $ 0.001) &Germany & 0.18& (0.005) \\ 
  Energy & 0.83 & ($< $ 0.001) &Great Britain & 0.38 & (0.003)\\ 
  Financials & 0.75 &  ($< $ 0.001) &Greece & 1.00 & ($< $ 0.001)\\ 
  Health Care & 0.55 &($< $ 0.001) & Italy & 0.25 & ($< $ 0.001)\\ 
  Industrials & 0.33 &($< $ 0.001) & Netherlands & 0.10 & (0.016)\\ 
  Information Technology & 0.93 & ($< $ 0.001) &Spain & 0.23 & ($< $ 0.001)\\ 
  Materials & 0.47 &($< $ 0.001) & Sweden & 0.41 & ($< $ 0.001)\\ 
  Telecommunications Services & 0.53 & ($< $ 0.001) & Switzerland & 0.11 & (0.007)\\ 
  Utilities & 0.77 & ($< $ 0.001) &  &  \\ 
   \hline
\label{tab:puritybefore}
\end{tabular}
}
\subfloat{
\begin{tabular}{lrl}
\hline
\multicolumn{3}{c}{After Transformation}\\
  \hline
 Country & Purity & \\ 
  \hline
 France  & 0.26 & ($<$ 0.001) \\ 
 Germany & 0.40& ($<$ 0.001) \\ 
   Great Britain & 0.75 & ($<$ 0.001) \\ 
  Greece & 1.00& ($<$ 0.001) \\ 
   Italy & 0.32& ($<$ 0.001) \\ 
 Netherlands & 0.16 & (0.002) \\ 
  Spain  & 0.93 & ($<$ 0.001) \\ 
 Sweden & 0.93 & ($<$ 0.001) \\ 
 Switzerland  & 0.15 & (0.004) \\ 
 & &\\
  \hline
\label{tab:purityafter}
\end{tabular}
}
\end{center}
\label{tab:purity}
\caption{The left hand table shows the purity of each sector and country before any adjustment is made to the data. The right hand table shows the purities after the adjustment discussed in Sec. \ref{sec:discovering} has been performed to remove the effects of sector/geography membership respectively. P-values are given in brackets.}
\end{table*}

Finally, we perform hierarchal clustering on the data using the algorithm discussed in Sec. \ref{sec:clustering}. Fig. \ref{fig:dendrosector} shows the resulting dendrogram with companies colored based on their sector membership, while Fig. \ref{fig:dendrocountry} shows the same dendrogram but with coloring based on geography. The clusters found by HC are as expected given the previous discussion, with the sector dendrogram showing more blocks of solid color (and hence more clustering structure) than the one which is colored based on geography, The Financial companies  in the bottom right corner (green in color, dark in greyscale) form an especially well defined cluster, but companies within the same sector tend to be clustered together throughout the dendrogram. In contrast the geography dendrogram is much more fragmented; it can be seen that neighboring companies are generally connected to those in the same country (as represented by many groups of 2-3 identical colors), however there is no large scale cluster formation due to the fact that companies are primarily split based on sector.  In order to assess this further, we will introduce a more quantitative measure to assess the extent to which the clustering is based on sector and geography. Then, we show how a transformation of the data can remove the effects of sector to make the geographical groupings more visible.

\subsection{Purity}
\label{sec:purity}

To quantify the extent that the clusters found in the data correspond to real sectorial/geographical groupings, the dendrogram produced by HC must be compared with the information about sectors and geography obtained from $S\&P$. However this is subtle because as we mentioned in Sec. \ref{sec:data} above, these groupings are rather broad and there may be several sub-sectors within each broad sector classification. A naive approach for assessing whether the clusters found by HC correspond to real sector/geography clusters would be to cut the dendrogram using some suitable entropy-based measure to form hard clusters, and then compare these to the $S\&P$ labels. Let $C_i$ be a cluster found in the data, and let $S_j$ be a collection of companies which share the same sector (or geographical) label according to $S\&P$. Then the overlap between $C_i$ and $S_j$ can be defined based on the number of companies they have in common .One possible measure for this is the Jaccard index \cite{Palla2007}:

$$\frac{|C_i \cap S_j|}{|C_i \cup S_j|}.$$

However this will not work for our purposes; suppose that the clustering algorithm splits the Financial companies into two subgroups, one containing banks and the other containing insurance companies. Both will have a low Jaccard index since neither corresponds exactly with the $S\&P$ list of Financial companies. To avoid this problem, we instead define a notion of purity which takes the hierarchal structure of the data into account, and which uses the dendrogram directly. Suppose we wish to quantify the extent to which companies in some sector $S_i$ are clustered together in the HC dendrogram. Let $c_i$ and $c_j$ be two arbitrary companies belonging to this sector. Recall from Sec. \ref{sec:clustering} that the HC algorithm hierarchally merges clusters until there is only one cluster remaining, which contains every company in the data set. There will therefore be a lowest point on the dendrogram where $c_i$ and $c_j$ are merged into the same cluster. Intuitively, this is the smallest cluster in the data containing both $c_i$ and $c_j$. The fraction of points in this cluster which belongs to the same sector as $c_i$ and $c_j$ is then computed. The procedure is then repeated for all pairs of companies in the sector, and the average value denotes the sector purity. A similar procedure can be used to measure the purity of each country. This procedure has previously been used by \cite{Heller2005} in a slightly different context.

In Table \ref{tab:puritybefore} (on the left), we show the purity scores for each of the sectors and countries in the data. It can be seen that these scores broadly reflect the qualitative features of the data we noted in the previous section, with the typical sector-based purity scores being substantially higher than those of the geography groupings. %The extreme values observed for Greece and Austria are partly a result of statistical noise; as Table \ref{tab:companysectorlist} shows there are only a very small number of companies from these countries present in the data, which makes it hard to get an accurate estimate of the clustering purity.

To aid interpretation of  Table \ref{tab:puritybefore}, it is important to assess whether the high purity scores represent real structure in the data, or whether they are purely a result of statistical noise. For this purpose, we use a measure of statistical significance similar to that of a permutation test \cite{Good2005}. Suppose that a particular sector or country contains M companies. In order to assess whether the observed purity score is significant, it should be compared to the distribution of purity scores that a random sample of M unrelated companies would have, since the latter represents the score that would be expected to arise through pure chance. The purity score distribution of a random sample can be estimated in the same manner as a permutation test \cite{Good2005}, which is commonly used in statistical applications. First, select a random sample of  $M$ companies from the data set, where each company has equal probability of being included in the sample. Next, compute the purity score of this sample; since the companies have been selected randomly, it is unlikely that the resulting score will be high. By repeating this procedure a large number of times using different randomly selected samples, the purity score distribution can be approximated. A p-value can then be computed for each of the purity scores in Table  \ref{tab:puritybefore}  based on the probability of a random sample of the same size having an equal or greater purity. As shown in the table, the resulting p-values are all very small, which suggests that the purity scores found for the sectors and countries given in Table  \ref{tab:puritybefore} do indeed correspond to genuine structures in the data. 

% \caption{Purity scores for each sector and company as a result of running the clustering algorithm over the whole period of data. Average purity scores across all sectors and countries are also given.}
%\label{tab:purityunadjusted}

\section{Discovering Hidden Clusters}
\label{sec:discovering}
In order to reveal the influence that geographical location has on the clustering structure, we propose adjusting each company's return series  to remove the effects of sector membership. If the effect that sector factors have on the returns is eliminated in a principled way, then the effect of geography should be clearer and less masked in the transformed data.

In previous studies of financial price movements, it has been found that a large percentage of the observed correlation structure can be explained by a single `market factor' which represents the market economy as a whole. In other words, much of the correlation between individual companies can be explained by the fact that each company is highly correlated with a global market factor. This phenomena has been found in many different financial data sets, often using Random Matrix Theory \cite{Coronnello2005}. Recently \cite{Borghesi2007} showed that if  the correlation matrix  is adjusted to remove the effect of the market factor, then the structure of the matrix becomes much less noisy -- after the correlation that each company has with the global factor has been removed, the remaining residual correlations allows clustering effects to be more clearly determined.

In \cite{Borghesi2007}, several methods for removing the effects of the global market factor are assessed. Each of these essentially involves subtracting the daily price returns of the market factor from the daily price returns of each individual company. The method which they found most successfully decouples the company correlations from the market factor, is as follows. First, the component of the returns associated with the market factor is estimated by a synthetic `pseudo-index' defined as the average returns of all the stocks in the portfolio:
\begin{equation}
R_{t} = \frac{1}{N} \sum_{i=1}^{N} r_{i,t}
\label{eqn:synthetic}
\end{equation}
The effect that this pseudo-index has on the returns of each stock can be removed by using regression. In a standard one variable linear regression model, the relationship between a set of dependent variables $\{Y\}_i$ and a set of independent variables $\{X\}_i$ is modeled as 
$$Y_i = \alpha + \beta X_i + \epsilon_i, \quad \epsilon \sim F$$
where $F$ is some symmetric distribution with mean 0. In this model, the expected value of $Y_i$ is $\beta_0 + \beta_1 X_i$, while each $\epsilon_i$ represents the residual variation in $Y_i$ which is not explained by $X_i$. The values of $\alpha$ and $\beta$ can be estimated using standard ordinary least squares.

If the returns of the pseudo-index are regressed on the returns of each stock,  this leads to the model:
\begin{equation}
r_{i,t} = \alpha_i + \beta_i R_{t} + \epsilon_{i,t},
\label{eqn:ols}
\end{equation}
where $r_{i,t}$ denotes the return of the $i^{th}$ stock on day $t$. The residuals $\epsilon_{i,t}$ then represents the components of the returns of stock $i$ which are not explained by the market factor. It is shown in  \cite{Borghesi2007}  that if each $r_{i,t}$ is replaced with $\epsilon_{i,t}$ when it comes to computing the correlations in Eqn. \ref{eqn:correlation} used for clustering, then the results tend to be less noisy since the correlation which each stock has with the market factor is removed. Note that it would be  also possible to perform the same adjustment by regressing the returns of each company onto those of a real stock index such as the $S\&P350$ itself, so that $R_{t}$ above is replaced with the daily returns of this index. However we choose to use the pseudo-index from Eqn. \ref{eqn:synthetic} instead, since \cite{Borghesi2007}  found via experimental analysis that using such a pseudo-index is more effective at decorrelating the companies than using a real index.

We propose using a similar approach to remove the effects of  sector  membership. Rather than constructing a pseudo-index to represent the market as a whole, we instead construct a separate pseudo-index for each sector separately, by averaging the returns of only the companies in that sector. Then, each company is regressed against the pseudo-index corresponding to the sector  it belongs to. In this way, the sector influence on each company is removed. Suppose that the $j^{th}$ GICS industrial sector $S^j$ contains the $M$ companies with return vectors $r_{(1)},\ldots,r_{(M)}$. The corresponding sector pseudo-index is then:
\begin{equation}
S^j_{t} = \frac{1}{M} \sum_{i=1}^{M} r_{(i),t}.
\label{eqn:synthetic2}
\end{equation}

Each company then has the component of its returns vector associated with its sector membership being removed, by being regressed against the corresponding sector index:
\begin{equation}
r_{(i),t} = \alpha_{(i)} + \beta_{(i)} S^j_{t} + \epsilon_{(i),t}.
\label{eqn:ols2}
\end{equation}

\begin{figure}[]
  \centering
  %\subfloat[Colored by sector]{\label{fig:mdssectoradjusted}\includegraphics[width=0.25\textwidth]{mdssectoradjusted.pdf}}
  \subfloat[ ]{\label{fig:mdssectoradjusted}\includegraphics[width=0.25\textwidth]{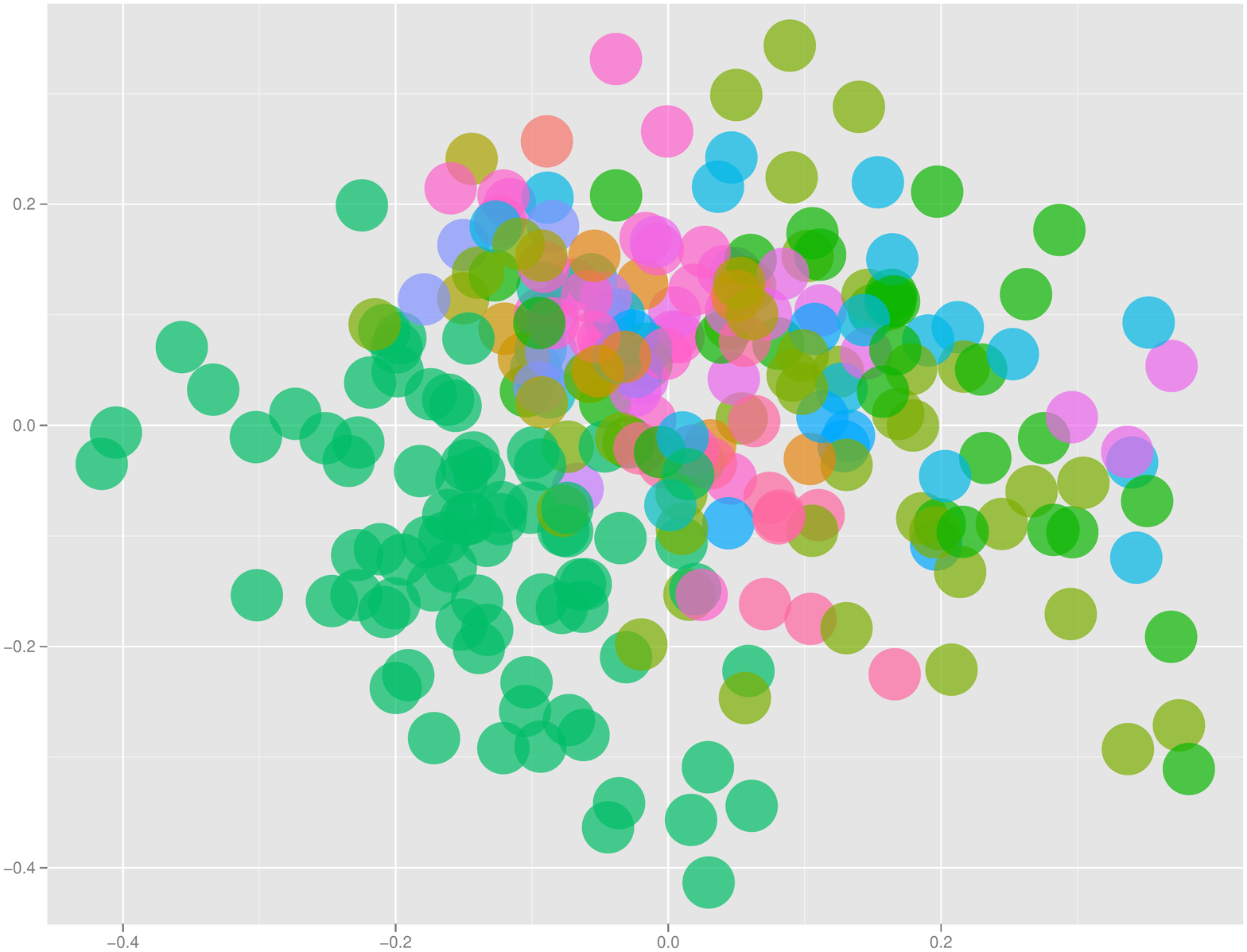}} 
  %\subfloat[Colored by sector]{\label{fig:dendrosectoradjusted}\includegraphics[width=0.25\textwidth]{dendrosectoradjusted.png}}
  \subfloat[  ]{\label{fig:dendrosectoradjusted}\includegraphics[width=0.25\textwidth]{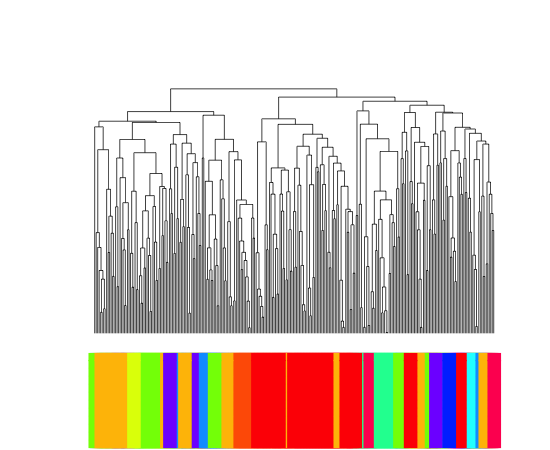}} 
%  \subfloat[Only industrials]{\label{fig:mdsonesector}\includegraphics[width=0.25\textwidth]{mdssector.pdf}}     
  %ONE MORE SUBFLOAT SHOWING CLUSTERING ALGORIHTM OUTPUT
\caption{MDS projection and dendrogram of the data after adjustment to remove the effects of sector membership, with companies colored based on their geographical location. It can be seen that removing the effect of sector membership gives a much clearer picture of how geographical location affects clustering.}
  \label{fig:mds3}
\end{figure}

Then, when it comes to computing the correlation matrix from Eqn. \ref{eqn:correlation} used for clustering, the returns of each company are replaced with the residual $\epsilon_{(i),t}$, representing the component of variance which is not associated with sector membership. As the effect of sector membership will then be largely removed from the data, the effects of geographical location should more clearly show up in the clustering structure.

However a problem occurs when using the above regression technique. Because we are considering each sector separately, there will only be a small number of companies which contribute to each sector pseudo-index $S^j$. As financial price return series  tend to be heavy-tailed and non-Gaussian \cite{Liu1999}, there is a risk that an extreme price movement in the returns of one company will drastically skew the results. We therefore instead use techniques from the field of robust regression analysis. First, rather than defining the pseudo-index as the mean of the $r_{(i),t}$ returns for each stock in the sector, we instead define it as the median. It is well known that the sample median is much less sensitive to outliers than the sample mean. We will now write $S^j_t$ for the resulting median-based pseudo-index for sector $j$ on day $t$.

Although this leads to more representative pseudo-indexes, if the regression coefficients $\alpha_i$ and $\beta_i$ are estimated via ordinary least squares, they will have a high variance if the residuals $\epsilon_{(i),t}$ come from a non-Gaussian heavy tailed distribution. We therefore instead use the Theil-Sen estimator to estimate each $\alpha_i$ and $\beta_i$ rather than the standard least squares estimator. The Theil-Sen approach is widely used when performing regression in situation where there may be extreme values, since it is more robust against outliers than the ordinary least squares \cite{Sen1968}. The Theil-Sen estimates can be computed as follows: for each company $i$ in sector $j$, write the data as the ordered pairs $(S^j_t,  r_{i,t})$ with $t=1,2,\ldots,T$. The slope estimator $\hat{\beta}_i$ is then defined as the median of all pairwise slopes between these pairs, i.e.:
$$\hat{\beta}_i = \mathrm{median}\left( \frac{r_{i,m} - r_{i,n}} {S^j_m - S^j_n} \right), \quad m \not= n,$$
where $m$ and $n$ run over all ordered pairs. Given this estimate, the intercept estimator $\hat{\alpha}_i$ is defined as the median of the $r_{i,t} - \hat{\beta}_i S^j_t$ values. As this estimator is defined in terms of medians, it is highly robust to outliers in the data and will give a better fit to the data than using the standard least squares approach.

After the price returns of each company have been transformed to remove the effects of sector membership according to Eqn. \ref{eqn:ols2} with the $\alpha_i$ and $\beta_i$ coefficients estimated using the Theil-Sen method, the clustering algorithm was rerun on the data. Fig. \ref{fig:mdssectoradjusted} shows a MDS plot of the companies colored by country, after the effect of sector membership has been removed. Compared to Figure \ref{fig:mdsgeography} from Sec. \ref{sec:static} the geographical clusters are now much more distinct, and there are several clear clusters. The green (dark in grayscale) cluster in the bottom left, which represents the countries located in the United Kingdom, make an especially distinct group, as do the brown Swedish companies in the upper right. Fig. \ref{fig:dendrosectoradjusted} shows the dendrogram which is produced by running the hierarchal clustering algorithm on this adjusted data; as expected, the geographical clusters found are much stronger than before, with much thicker bands of color.

In order to quantitatively assess the impact which the removal of sector membership has on the purity of the geographical clusters found by HC, Table \ref{tab:purityafter}, gives the new purity associated with each country. These scores are now much higher than those in  Table\ref{tab:puritybefore}  before the regression adjustment was made, with almost every country showing a much higher purity, particularly Finland, Germany, the United Kingdom, Portugal, Spain and Sweden.  This analysis shows that contrary to first impressions, geography plays a strong role in determining the clustering structure. It is only the fact that sector membership has an even stronger influence which masks this.

%adjustements
\section{Dynamic Analysis}

We now investigate how the clustering purity of each country changes over time. This is important since there is no prior reason to believe that the effect of geography will be static, particularly with the recent financial crisis and sovereign debt concerns. As discussed in the Introduction, this has potential implications for risk management since assessing the risk of a portfolio should take into account potential changes in the dynamic correlation structure, and the potential for companies to cluster based on different factors at different times.

\begin{figure*}[]
  \centering
  \subfloat[]{\label{fig:dynamiccountries1}\includegraphics[width=0.4\textwidth]{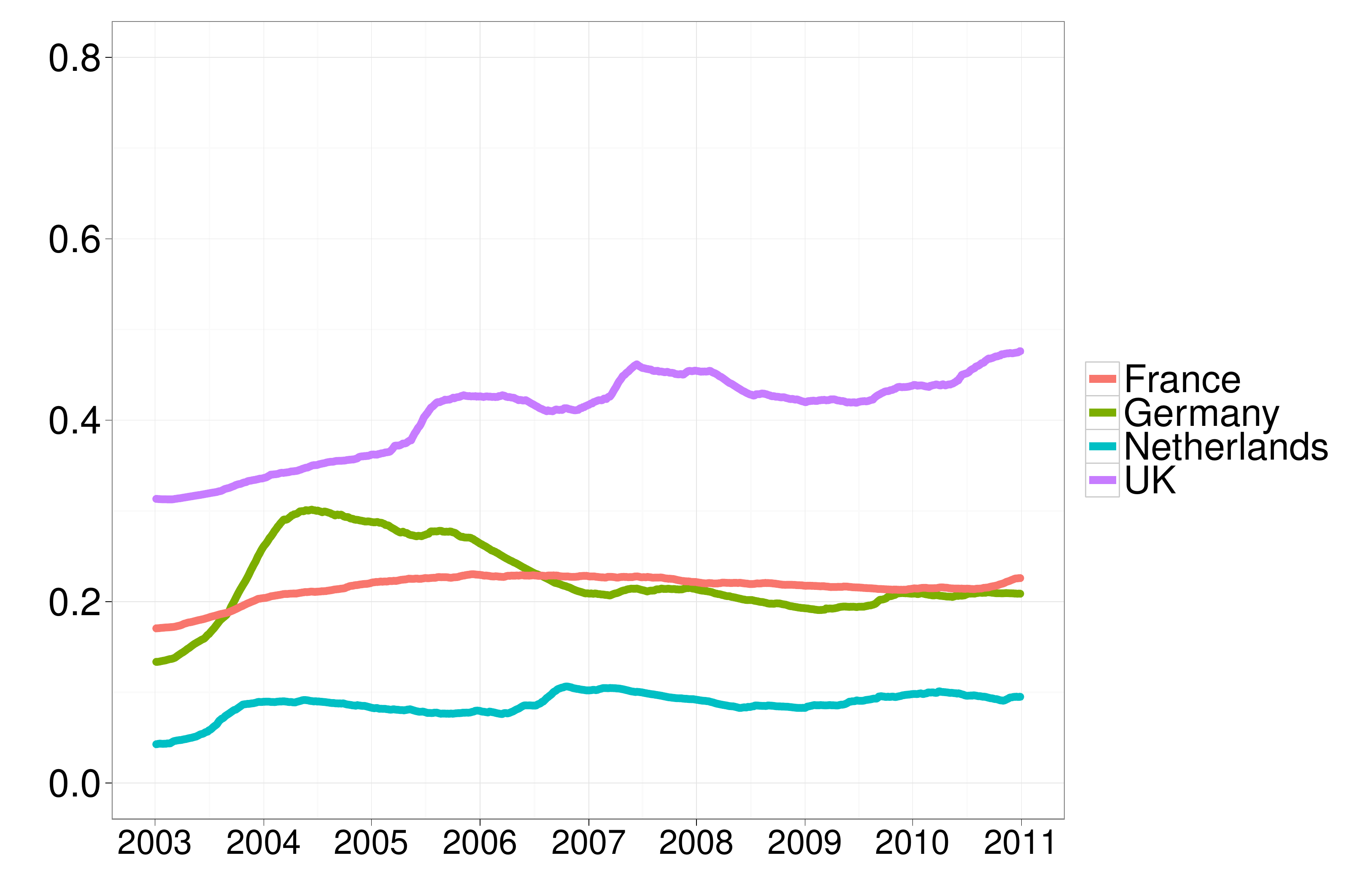}}
  \subfloat[]{\label{fig:dynamiccountries2}\includegraphics[width=0.4\textwidth]{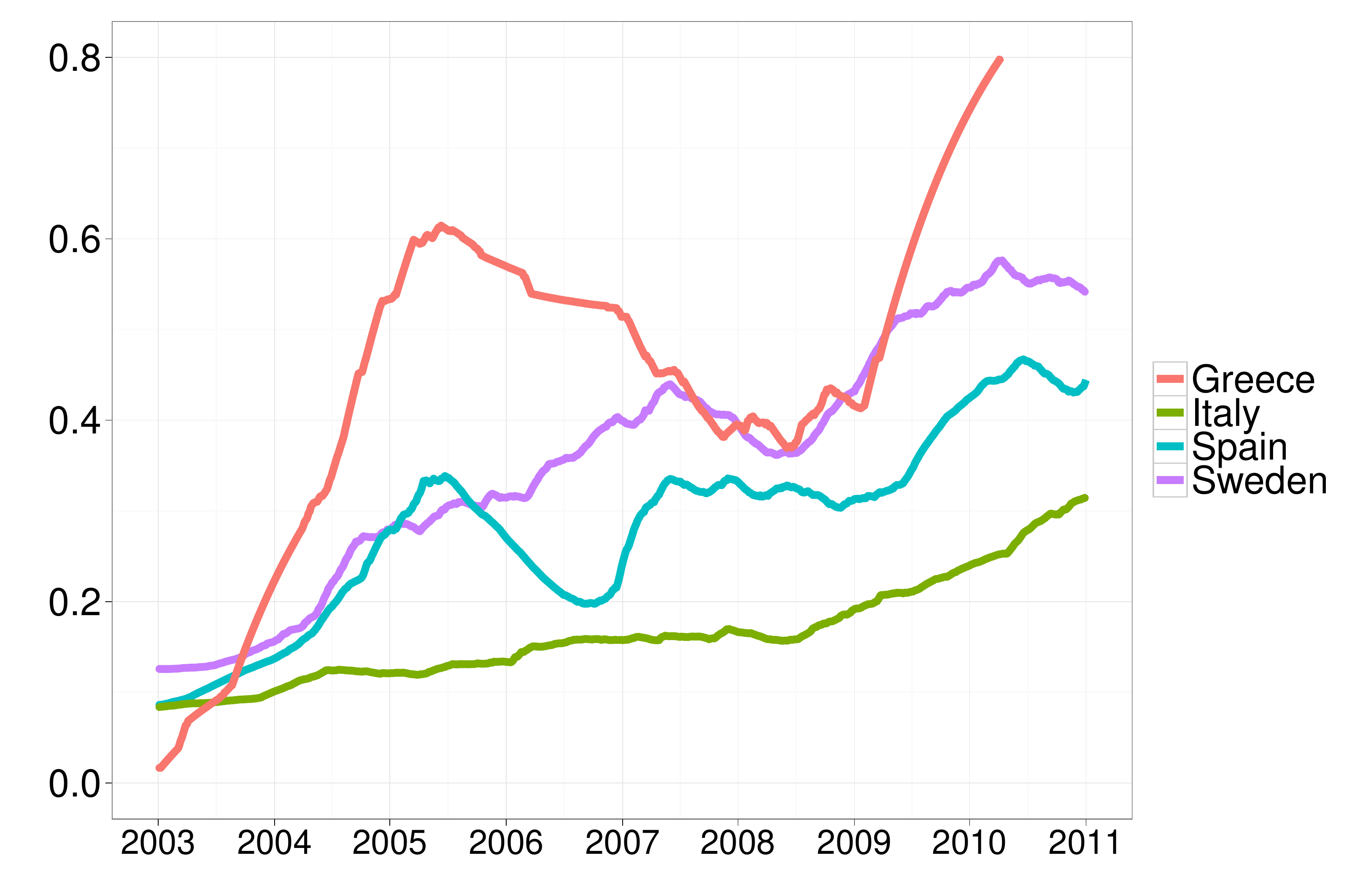}} \\
  \subfloat[]{\label{fig:dynamicunadjusted1}\includegraphics[width=0.4\textwidth]{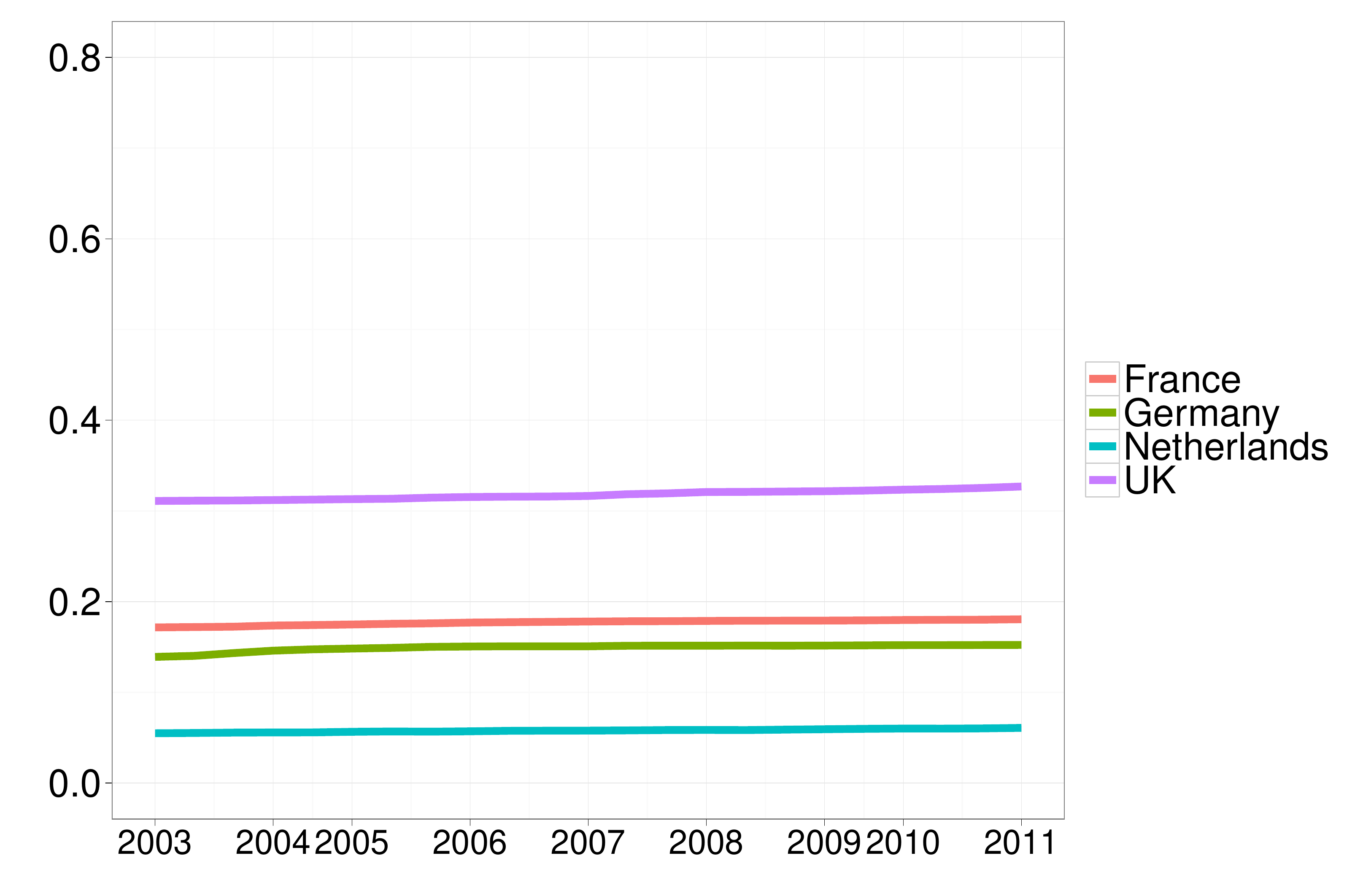}}
  \subfloat[]{\label{fig:dynamicunadjusted2}\includegraphics[width=0.4\textwidth]{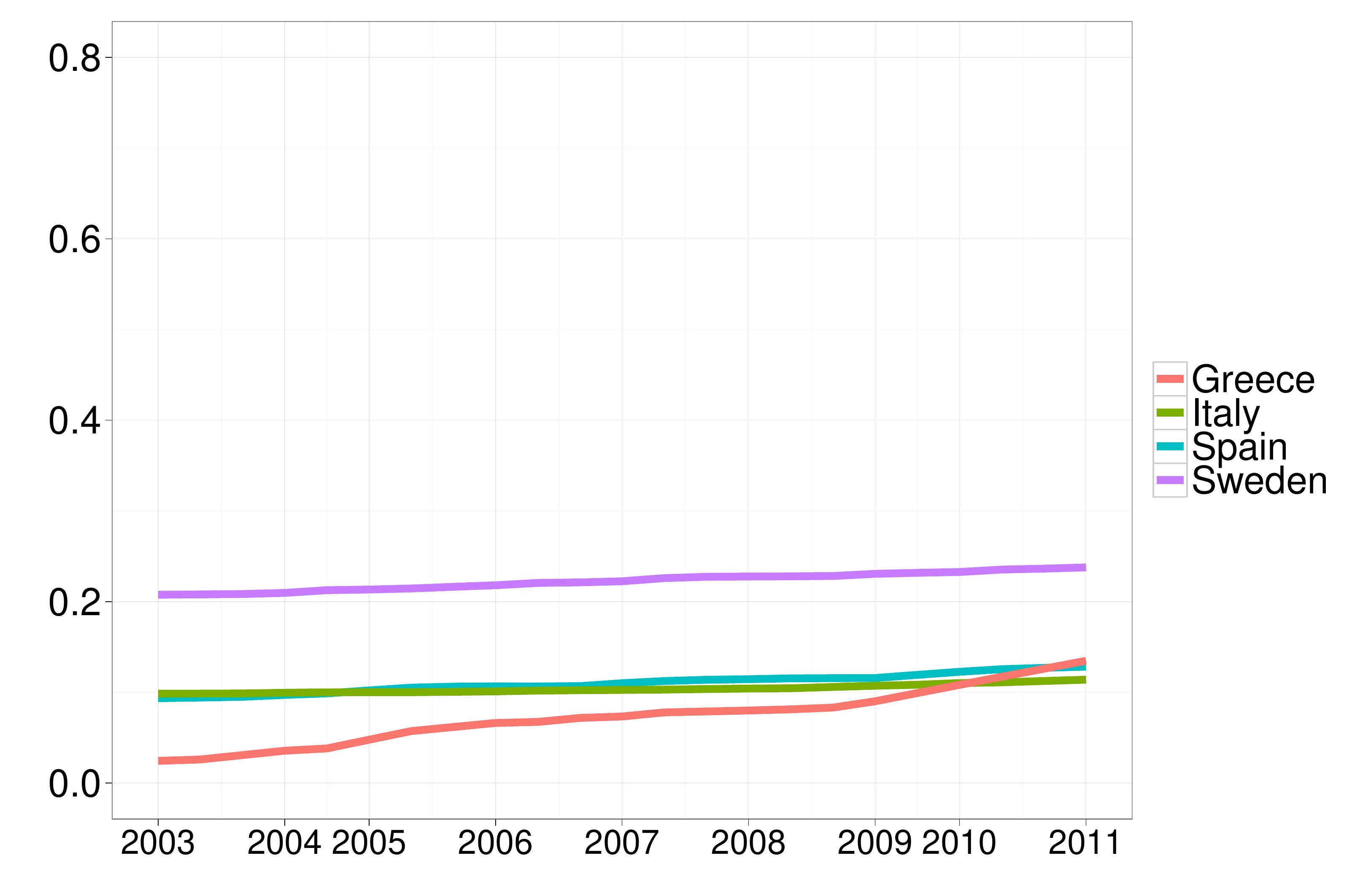}} 
\caption{Figs. (a) and (b) show how the cluster purity scores for each country change over time after the influence of sector membership has been removed via the technique in Sec. \ref{sec:discovering}. The increased purity scores of the most crisis-hit countries after 2008-2009 are very prominent .Figs. (c) and (d) show the same purity scores computed on the data that has not had sector membership removed. The dynamic structure is clearly masked in these, and cannot be seen.}
  \label{fig:dynamic}
\end{figure*}

 In order to study how the clustering structure evolves over time, we must allow the correlation $\rho_{ij}$ between companies $i$ and $j$ to be time-varying, rather than fixed over the whole period as it was in Eqn. \ref{eqn:correlation} in the static case. Given an appropriately defined time-varying correlation $\rho_{ij,t}$, a dynamic distance measure between the companies can be defined in a similar way as in the static case:
\begin{equation}
W_{ij,t} = \sqrt{2 (1-\rho_{ij,t})}.
\end{equation}

In order for  $\rho_{ij,t}$  to accurately estimate the true (dynamic) correlation at time $t$, it must be defined in a way which gives most weight to the recent stock prices at $t$ and less weight to older prices. The  usual way to implement this \cite{Conlon2009,Fenn2011} is via the notion of a sliding window where the correlation at time $t$ is computed using only the most recent $w$ observations, for some choice of the window length $w$. However we feel that this weighting is slightly unrealistic;, with the last $w$ observations receiving equal weights and then an abrupt drop-off where older observations receive no weight at all. Therefore we instead use exponential forgetting which allows the weight assigned to older observations to decay more smoothly. Let $\lambda \in [0,1]$ denote a forgetting factor, where a value closer to $1$ results in more weight  being given to recent observations. The time-varying variance of company $i$ can then be defined as: 
$$\sigma^2_{i,t} = (1-\lambda)\sigma^2_{i,t-1} + \lambda \tilde{r}_{i,t}^2, \quad \sigma^2_{i,0} = 0  $$
and the time-varying correlation between companies $i$ and $j$ can be similarly defined:
$$\rho_{ij,t} = (1-\lambda)\rho_{ij,t} + \lambda \frac{\tilde{r}_{i,t} \tilde{r}_{j,t}} {\sigma_{i,t} \sigma_{j,t}}, \quad \rho_{ij,0} = 0. $$
This implements exponential smoothing through a weighted moving average which assigns higher weights to more recent observations, with $\lambda$ controlling the rate at which older observations are forgotten. We used the value $\lambda=0.01$ in the following analysis in order to reduce the effect of short-term fluctuation, particularly in the countries where data on only a small number of countries is available. The distance matrix can then be defined at each time step as $W_{ij,t}$ above, which is used for computing the clustering using HC

Note also that we have marked the returns  $\tilde{r}$ with a tilde to emphasize that we are using the data which has had sector-influence removed via the techniques introduced in the previous section -- as will be seen below, this makes the dynamic effect of geographical much more visible.

%note that we used the sector adjusted data
For each of the 2225 days in the 8 year period, we generated the corresponding $W_{ij,t}$ matrix. The HC clustering algorithm was run over each matrix, and the purity scores of the resulting clusters were computed on each day. Figures \ref{fig:dynamiccountries1} and \ref{fig:dynamiccountries2} show how the clustering purity for several countries changes over time. These plots show that that in the countries most affected by the financial crisis, namely Spain, Greece and Italy, there is a very noticeable increase in the clustering purity after $2008-2009$, corresponding to the beginning of the European sovereign debt crisis. This is broadly consistent with other literature \cite{Hartmann2004, Sandoval2012} which suggests that financial assets become more correlated in times of crisis. However our analysis clearly shows that this does not affect all countries equally, since Great Britain, Germany, France and the Netherlands do not show such a sharp increase, which suggests companies in these countries are not clustering together to a increased extent. Interestingly, companies in  Sweden also displayed substantially higher clustering in the aftermath of the financial crisis. This may initially seem surprising since this was one of the countries most unaffected by  global instability. However, it is well known that in times of market uncertainty, investors prefer to invest in assets which are considered relatively safe in order to hedge against wider risk. This phenomena is known as the ``flight to quality'' \cite{Caballero2008} and is a likely explanation of the increased correlation between Swedish assets,  as risk-averse investors flocked to invest in these safe-havens. %This is consistent with recent studies on the reaction of safe Scandinavian securities to the crisis CITE.

For reference, Figures \ref{fig:dynamicunadjusted1} and \ref{fig:dynamicunadjusted1} show the results of the same analysis carried out on the raw data which has not had the influence of sector membership removed as in Sec. \ref{sec:discovering}. In this case there are no obvious geographical patterns in the data, and the increase in clustering in the crisis-hit countries after 2008 cannot be seen. This reenforces the results previously shown in Table \ref{tab:purityafter}  that it is important to remove sector effects before studying geography; if this is not done then the country effects are invisible.

\section{Concluding Remarks}

In this work we have studied the effect that multiple correlation factors can have on the clustering structure of a financial portfolio. Unlike previous studies which have considered the effect of sector membership and geographical location separately, we studied both simultaneously and have shown that the interaction between these factors can be complex, and not easily handled by existing techniques for identifying overlapping communities.

In order to reveal the time-varying dynamics of the clustering structure, we proposed a method for adjusting the observed correlation matrices to remove the effects of both sector membership and geography separately. After making this adjustment, the effect that both factors have on the clustering becomes much clearer. This allows the  clustering structure to be studied over time. Our analysis reveals that between the years of 2003 and 2008, companies located in the same country did not tend to cluster together. However after the 2008 financial crisis this pattern changed, and geography started to become an important determinant of the clustering. This is especially the case in the countries which were hit hardest by the crisis, such as Spain and Italy.  

\bibliographystyle{apsrev4-1}
\bibliography{jabref}	

\end{document}